\documentclass[a4paper,11pt]{article}
\usepackage{jheppub} 
\usepackage{lineno}
\usepackage{amssymb,color,graphics,epsfig}
\usepackage{multirow}
\usepackage{amsmath}
\usepackage{mathrsfs}
\usepackage{hyperref}

\usepackage[shadow,textwidth=2.6cm]{todonotes}
\usepackage{ifthen}
\reversemarginpar
\newcounter{todocounter}
\definecolor{aqua}{rgb}{0.0, 1.0, 1.0}
\colorlet{ypcolor}{green!40!white}

\newcommand{\ypinline}[2][]{
	
	\ifthenelse { \equal {#1} {} }
	
	{ \def\temp {#2} }  
	
	{ \def\temp {#1} }   
	
	\refstepcounter{todocounter}\todo[color=ypcolor,inline,caption={\textbf{\thetodocounter. YP} \temp}]{\textbf{\thetodocounter. YP:} #2}{}}

\colorlet{hpjcolor}{blue!40!white}

\newcommand{\hpjinline}[2][]{
	
	\ifthenelse { \equal {#1} {} }
	
	{ \def\temp {#2} }  
	
	{ \def\temp {#1} }   
	
	\refstepcounter{todocounter}\todo[color=hpjcolor,inline,caption={\textbf{\thetodocounter. HPJ} \temp}]{\textbf{\thetodocounter. HPJ:} #2}{}}

\usepackage{amsfonts}

\newcommand{\be}{\begin{equation}}
	\newcommand{\ee}{\end{equation}}
\newcommand{\bea}{\setlength\arraycolsep{2pt} \begin{eqnarray}}
	\newcommand{\eea}{\end{eqnarray}}
\newcommand{\nn}{\nonumber}

\def\ft#1#2{{\textstyle{\frac{\scriptstyle #1}{\scriptstyle #2} } }}

\def\0{{\sst{(0)}}}
\def\1{{\sst{(1)}}}
\def\2{{\sst{(2)}}}
\def\3{{\sst{(3)}}}
\def\4{{\sst{(4)}}}
\def\5{{\sst{(5)}}}
\def\6{{\sst{(6)}}}
\def\7{{\sst{(7)}}}
\def\8{{\sst{(8)}}}
\def\sst#1{{\scriptscriptstyle #1}}

\def\a{\alpha}
\def\b{\beta}
\def\g{\gamma}

\def\d{\delta}

\def\k{\kappa}
\def\l{\lambda}
\def\L{\Lambda}
\def\m{\mu}
\def\n{\nu}
\def\r{\rho}
\def\s{\sigma}

\def\o{\omega}

\def\wt{\widetilde}
\def\bb{\boldsymbol{b}}
\def\bg{\boldsymbol{g}}

\title{\boldmath  A holographic model of magnetohydrodynamics with fortuitous SO(3) symmetry}

\author[a]{Yanqi Wang}
\author[a]{Peng-Ju Hu}
\author[a,b]{Yi Pang}

\affiliation[a]{Center for Joint Quantum Studies and Department of Physics,\\
	School of Science, Tianjin University, Tianjin 300350, China}
\affiliation[b]{Peng Huanwu Center for Fundamental Theory, Hefei, Anhui 230026, China}

\emailAdd{wangyanqi0@gmail.com}
\emailAdd{pengjuhu@tju.edu.cn}
\emailAdd{pangyi1@tju.edu.cn}

\abstract{
	We study  magnetohydrodynamics using holography. The gravity model is closely related to the STU supergravity in five dimensions and admits an analytical black brane solution carrying the conserved charge dual to the magnetic 1-form symmetry of the magnetohydrodynamic system. The black brane solution features a fortuitous SO(3) symmetry, providing a new symmetry principle for describing the magnetohydrodynamics. Since the bulk theory contains multiple 2-form gauge fields, the resistivity becomes matrix-valued. We find that the antisymmetric part of the resistivity  matrix exhibits novel features depending on the UV cut-off of the theory. We also compute the shear and bulk viscosities and find that the bulk viscosity is proportional to the shear viscosity. Remarkably, the proportionality constant is exactly what is required for conformality, despite the zeroth-order energy-momentum tensor not being trace-free.

}
\begin{document}
	
	\maketitle
	
	\flushbottom
	\section{Introduction}
	Magnetohydrodynamics (MHD) is essential in diverse fields, including  plasma physics \cite{Bellan_2006,Goedbloed_Poedts_2004,Goedbloed_Keppens_Poedts_2010}, quantum chromodynamics (QCD) \cite{Li:2017tgi,An:2021wof} and superfluidity \cite{Armas:2018atq,Armas:2018zbe}. Understanding transport coefficients such as viscosity and resistivity within MHD is crucial for both theoretical research \cite{Grozdanov:2016tdf,Grozdanov:2017kyl} and experimental investigations \cite{sisan2004experimental,highcock2010transport,cabanes2014turbulence}. However, the inherent complexity of magnetohydrodynamics (MHD) \cite{Balbus:1998ja,Davidson_2001,Brandenburg:2004jv} poses significant challenges for traditional analytical methods and numerical simulations when calculating these transport coefficients \cite{Hopkins:2015epa}. An alternative approach is based on the gauge/gravity duality,  which links a certain gravitational theory to fluid dynamics and offers a novel perspective for calculating transport coefficients \cite{Policastro:2001yc,Buchel:2003tz,Kovtun:2004de,Son:2007vk,Erdmenger:2008rm,Banerjee:2008th,Romatschke:2009im,Jeon:2015dfa} ( see \cite{Baggioli:2023yvc,Ge:2023yom} for recent developments). Within this framework, transport coefficients can be determined from perturbations of fields in the bulk gravitational theory. A notable result of applying gauge/gravity duality to hydrodynamics is a universal lower bound on the ratio of shear viscosity $\eta$ to entropy density $s$ \cite{Kovtun:2004de}, specifically $\eta/s \geq \frac{1}{4\pi}$, in the absence of an external magnetic field and non-minimal couplings. Interestingly, previous research \cite{Grozdanov:2016tdf, Grozdanov:2017kyl, Rangamani:2009xk} has shown that transport coefficients change significantly in the presence of strong magnetic fields, offering clear theoretical predictions for future experimental tests.

	In magnetohydrodynamics, a system is characterized by the  energy-momentum tensor $T_{\mu\nu}$, and the 2-form current $J^{\mu\nu}=\ft12\epsilon_{\m\n\r\l}F^{\r\l}$, where is $F_{\m\n}$ is the field strength of the electromagnetic gauge field. In the long-wavelength limit, the fundamental equations of motion in magnetohydrodynamics are the conservation equations \cite{Grozdanov:2016tdf}
	\be
	\nabla_{\mu}T^{\mu\nu}=H_{\ \mu\r}^{\nu}J^{\mu\r}\ ,\ \ \nabla_{\mu}J^{\mu\nu}=0\ ,
	\ee
	where $H_{(3)}=db_{(2)}$ with  $b_{\mu\nu}$ representing a 2-form background gauge field coupled to the 2-form current $J^{\mu\nu}$. As usual,  the energy-momentum tensor $T_{\mu\nu}$ is coupled to the background metric. In gauge/gravity duality, it has been suggested that \cite{Grozdanov:2016tdf, Hofman:2017vwr,Grozdanov:2017kyl} the magnetohydrodynamics in 3+1 dimensions can be described by a gravity model with a negative cosmological constant in 4+1 dimensions, consisting of a metric $G_{MN}$  and a 2-form gauge fields $B_{MN}$ where $M,N=0\cdots4$. The action of the gravity model takes the form \cite{Hofman:2017vwr, Grozdanov:2017kyl}
	\be
	S_{5}= \frac{1}{2\kappa^{2}} \int d^{5}x \sqrt{-g} \big( R+12- \frac{1}{12} H_{MNP} H^{MNP} \big)+S_{\rm bndy},\quad H_{MNP}=3\nabla_{[M} B_{NP]}\ ,
	\ee
	where $2\k^2=16\pi G_5$ and the AdS radius is set to 1 for notational simplicity. In \cite{Hofman:2017vwr, Grozdanov:2017kyl}, this model has been applied to
	study transport properties of hydrodynamics in which the anisotropy is introduced by the external magnetic field breaking the spatial SO(3) invariance to SO(2) invariance in directions orthogonal to the magnetic field.
	
	In this paper, we explore a holographic model of magnetohydrodynamics that displays anisotropy in all three spatial directions. The gravity model includes three 2-form fields $B_{\m\n}^a$ with $a=1,2,3$ representing different flavor indices. By applying the same boundary conditions for the 2-form gauge fields as in \cite{Grozdanov:2017kyl}, the gauge/gravity duality dictionary implies that there exist three dynamical U(1) gauge fields in the dual fluid. Accordingly, the conservation equations of the fluid
	are given by
	\be
	\nabla_{\n}T^{\m\n}=\sum_{a=1}^3 H^{a\m}{}_{\r\s}J^{a \r\s},\quad \nabla_{\n} J^{a \m\n}=0\ .
	\ee
	Although the number of dynamical U(1) gauge fields present in this toy model does not match that in the real physical systems, it might still emulate the real physical situation and provide valuable insights on its hydrodynamic properties. We also note that in string theory-inspired grand unified models, the presence of multiple U(1) gauge fields is common. In such cases, magnetohydrodynamics with multiple U(1) gauge fields could indeed describe the early universe. Using the standard holographic techniques, various transport coefficients can be derived from the perturbations of $G_{MN}$ and $B_{MN}$ about a background black brane solution asymptotic to AdS$_5$.
	Crucially, this model admits an analytical black brane solution, which significantly simplifies the computation of transport coefficients. In contrast, previous studies have predominantly relied on numerical solutions \cite{DHoker:2009ixq}. As we will discuss later, errors would have arisen if the analytical solution were not available. The black brane solution in the model above is, in fact, the electromagnetic dual of the magnetic black brane in the 5-dimensional STU model \cite{Donos:2011qt}. Aside from this advantage, the model also has several other notable features, listed below:
	\begin{enumerate}
		\item Unlike the single U(1) case \cite{Grozdanov:2017kyl, Bhattacharyya:2007vjd}, the background solution here exhibits a combined SO(3) symmetry that acts simultaneously on both the flavor indices and the spatial indices of the 2-form gauge fields (see \eqref{mb2} for details). It turns out that this symmetry provides a new organizational principle for various transport coefficients, resulting in many of them being equal, even though the continuous SO(3) symmetry in the spatial directions is broken. This fortuitous SO(3) symmetry also suggests defining the Hodge dual of each 2-form current with respect to the three spatial directions as $\widetilde{J}^{ab} = \frac{1}{2} \epsilon^{acd} J_{cd}^{b}$, which can be further decomposed into symmetric and antisymmetric parts: $\widetilde{J}^{ab} = \widetilde{J}^{(ab)} + \widetilde{J}^{[ab]}$.
		\item  The resistivity $r_{[ab]}$ associated with the antisymmetric part of the conserved current $\wt{J}^{[ab]}$ displays intricate behavior depending on whether an external magnetic field is present or not. When the magnetic field is turned off, $r_{[ab]}$ is inversely proportional to the temperature. However, in the presence of a non-zero magnetic field, $r_{[ab]}$ exhibits a discontinuity at a critical temperature that depends on the magnetic field and the UV cutoff of the system. Thanks to the availability of an analytical background solution, we are able to solve the relevant perturbations exactly, leading to the discovery of this novel resistivity behavior not observed in previous studies \cite{Hofman:2017vwr, Grozdanov:2017kyl}.
		
		\item We also find that in our model, all the shear viscosity coefficients are equal. As for the bulk viscosity coefficients, the diagonal and off-diagonal components are separately equal. We also observe that the bulk viscosity is proportional to the shear viscosity.
		In particular,  $\zeta_{11} = \frac{4}{3}\eta$ and $\zeta_{12}= -\frac{2}{3}\eta$ for all values of the magnetic field. In our parameterization of the energy-momentum tensor, these relations resemble that of the conformal fluid, although at the zeroth order, the conformality is broken by the external magnetic fields.
		
	\end{enumerate}

	This paper is organized as follows. In section \ref{section2}, we study the thermodynamic properties of the background solution and obtain the zeroth order energy-momentum tensor and the 2-form currents. In section \ref{section3}, we define various transport coefficients in the anisotropic magnetohydrodynamic system and establish the semi-positivity of these coefficients along with the Kubo formulae. In section \ref{section4}, we compute all the first order transport coefficients. We conclude with discussions in section \ref{section5}.
	
	\section{The background solution }\label{section2}
	In this section, we briefly review the AdS magnetic brane solution in the $D=5$ truncated STU model.
	By performing a Hodge duality on this solution, we obtain the background solution in the  gravitational model with three 2-form gauge fields.
	The truncated STU model is of the form
	\be
	S_{\rm ren}=\frac1{2\k^2}\int d^5x\sqrt{-g}\Big(R+12-\frac{1}{4}\sum_{a=1}^{3} F^{a}_{MN} F^{aMN}\Big)+S_{\rm bndy}\ ,
	\label{stutrunc}
	\ee
	which captures static purely electric or magnetic solutions with constant scalars in the full STU model.
	The complete boundary action including the Gibbons-Hawking term and counterterms is given by \cite{DHoker:2009ixq}
	\be
	S_{\rm{bndy}} =\frac{1}{2\kappa^{2}}\int_{\partial M_{}}\sqrt{-\gamma}d^{4}x\Big(2K-6-\frac12R+\frac{1}{4}\ln ({r_c\L})\sum_{i=1}^{3}F_{\m\n}^{a}F^{a \m\n}\Big)\ ,
	\ee
	where $\g_{\m\n}$ denotes the induced metric on the boundary located at $r=r_c$ and $K$ is the extrinsic curvature. As discussed in \cite{Grozdanov:2017kyl}, $\L$ corresponds to the UV cut-off energy scale of the dual theory.
	
	A salient property of the model is that it admits an analytical AdS magnetic black brane solution given below
	\cite{Donos:2011qt}
	\bea
	ds_5^2&=&-fdt^2+\frac{dr^2}{f}+r^2(dx_1^2+dx_2^2+dx_3^2),\quad
	f=r^2-\frac{r^4_+}{r^2}+\frac{B^2}{2r^2}\ln\frac{r_+}{r}\ ,
	\nn\\
	F^{(1)}&=& B dx_2\land dx_3,\quad F^{(2)}=B dx_3\land dx_1,\quad F^{(3)}=B dx_1\land dx_2\ ,
	\label{mb1}
	\eea
	which is invariant under the scaling transformation
	\bea
	r\rightarrow \lambda r,\quad t\rightarrow  t/\lambda,\quad
	x_i\rightarrow x_i/\lambda,\quad B \rightarrow\lambda^2 B,\quad r_+\rightarrow\lambda r_+\ .
	\label{ss}
	\eea
	One can check that for the field equations to be satisfied, the three U(1) gauge fields must carry the same amount of magnetic flux.
	The model \eqref{stutrunc} admits a dual formulation in terms of  2-form gauge fields $B^a_{MN}$ whose field strength $H^a_{MNP}$ is Hodge dual to $F^a_{MN}$ via $H^{a}_{MNP}=\ft1{2}\epsilon_{MNPRS}F^{a RS}$. In the dual formulation, the action becomes
	\bea
	\widetilde{S}_{\rm ren} &=&\frac{1}{2\kappa^{2}}\int d^{5}x\sqrt{-g}(R+12-\sum_{a=1}^{3}\frac{1}{12}H_{MNP}^{a}H^{a MNP})+\widetilde{S}_{\rm bndy}\ ,
	\nonumber \\
	\widetilde{S}_{\rm bndy} & =&\frac{1}{2\kappa^{2}}\int_{\partial M}d^{4}x\sqrt{-\gamma}\Big(2K-6-\frac{1}{2}R[\gamma]+\frac{1}{4}\ln ({r_c\L})\sum_{a=1}^{3}{H}_{MNP}^{a}n^{P}{H}^{a MNQ}n_{Q}\Big)\ ,
	\label{dualS}
	\eea
	where $n^P$ is the outward unit normal vector of the boundary defined at $r=r_c$.  In terms of the 2-form $B^{a}_{MN}$, the magnetic brane solution \eqref{mb1} becomes an electric brane solution charged under the 2-form gauge fields and takes the form
	\bea
	ds^{2} & =&\frac{1}{u}\left(-F(u)dt^{2}+dx_{1}^{2}+dx_{2}^{2}+dx_{3}^{2}\right)+\frac{du^{2}}{4u^{2}F(u)}\ ,\nonumber \\
	F(u) & =& 1-\frac{u^{2}}{u_{+}^{2}}+\frac{B^2u^2}{4}\text{ln}\frac{u}{u_{+}}\ ,\quad
	H^{a}=\frac{B}{2u}du\land dt\land dx_a \ ,\quad a=1,2,3\ ,
	\label{mb2}
	\eea
	where we introduced the new coordinate $u:=1/r^2$, in terms of which the black brane
	horizon and the IR cutoff surface are now located at $u_{+}=1/r_{+}^{2}$ and $u_c=1/r_c^2$ respectively. The AdS boundary corresponds to $u_c\rightarrow 0$. It is noteworthy that
	the nonvanishing component of each $B^a_{MN}$ is of the form
	\be
	B^a_{t x_i}=\frac{B}2\delta^a_i \ln u  \ ,
	\ee
	which in invariant under a combined SO(3) rotation acting simultaneously on the flavor index $a$ and the spatial index $i$.  This symmetry was also noticed by \cite{Meiring:2023wwi} in a different context.

	In the long wavelength limit, the magnetohydrodynamics is described by the energy-momentum tensor and the 2-form current \cite{Hofman:2017vwr} which is Hodge dual to the field strength of the boundary Maxwell gauge field. In the framework of gauge/gravity duality, they can be computed using holographic renormalization \cite{deHaro:2000vlm, Grozdanov:2016tdf,Hofman:2017vwr}. The form of the magnetic brane solution \eqref{mb2} suggests the following general behaviors of the metric and the 2-form gauge fields near the AdS boundary
	\bea
	ds^2&=&\frac{du^2}{4u^2}+\frac1u\big(g^{(0)}_{\m\n}(x)+u g^{(1)}_{\m\n}(x)+{\cal O}(u^2)+\cdots\big)dx^\m dx^\n
	\nn\\
	B^a_{u\m}&=&0,\quad B^a_{\m\n}=B^{a(0)}_{\m\n}+\ln u\, B^{a(1)}_{\m\n}+\cdots\ .
	\label{FG}
	\eea
	Based on the expressions above, varying the metric and 2-form gauge fields in the action \eqref{dualS} leads to
	\be
	\delta \widetilde{S}_{\rm ren}=\int_{\partial M} d^{4}x\sqrt{-g^{(0)}}\big(\frac{1}{2} T^{\mu\nu} \delta g_{\mu\nu}^{(0)}+\sum_{a=1}^{3}J^{a \mu\nu}\delta b_{\mu\nu}^{a}\big)\ ,
	\label{vs}
	\ee
	where $\delta b^a_{\mu\nu}:=\delta B_{\mu\nu}^{a(0)}+2\ln (\Lambda L)\delta B_{\mu\nu}^{a (1)}$.  Setting $b^a_{\mu\nu}=0$ is equivalent to imposing the mixed boundary condition on each $B_{\m\n}^a$ which further implies the existence of three dynamical Maxwell gauge fields in the dual field theory, generalizing the scenario studied in \cite{Grozdanov:2017kyl, DeWolfe:2020uzb} with a single U(1) gauge field in the dual field theory. From \eqref{vs}, we read off the holographic energy-momentum tensor and the 2-form currents as follows
	\bea
	T_{\mu\nu} & =&-\lim_{u_c\rightarrow0}\frac{1}{\kappa^{2}\,u_c}\big(K_{\mu\nu}-K\gamma_{\mu\nu}+3\gamma_{\mu\nu}-\frac{1}{2}G_{\mu\nu}[\gamma]+\frac{1}{2}\text{ln}(\L/ \sqrt{u_c})\sum_{a=1}^{3}T[\mathcal{H}^{a}]_{\mu\nu}\big)\Big|_{u=u_c}\ ,
	\nonumber \\
	J_{\mu\nu}^{a}&=&-\frac{1}{4\kappa^{2}}\lim_{u_c\rightarrow0}\mathcal{H}_{\mu\nu}^{a}\Big|_{u=u_c}\ ,
	\label{HTHJ2}
	\eea
	where $\mathcal{H}_{\mu\nu}^{a}=-2uH_{u\mu\nu}^{a}$ and $\ T^{a}[\mathcal{H}]_{\mu\nu}=\mathcal{H}_{\mu\sigma}^{a}\mathcal{H}_{\nu}^{a\sigma}-\frac{1}{4}\gamma_{\mu\nu}\mathcal{H}^{a}_{\r\s}H^{a\r\s}$.
	
	Evaluating the 2-form currents on the solution \eqref{mb2}, we find that the non-zero component of each 2-form current is given by
	\be
	J^a_{tx_i}=\frac{B}{4\k^2}\delta^a_i\ .
	\ee
	
	We now turn to the evaluation of the holographic energy-momentum tensor on the electric black brane solution \eqref{mb2}. First of all, $T_{tt}$ gives rise to
	the energy density $\varepsilon$
	\be
	\varepsilon=\frac{3\left(B^{2}u_{+}^{2}\log(u_{+}\Lambda^{-2})+4\right)}{8\kappa^{2}u_{+}^{2}}\ .
	\ee
	In magnetohydrodynamics, the diagonal component of the energy-momentum tensor in each spatial direction receives contributions from both the pressure and the magnetic flux \cite{Grozdanov:2017kyl, Grozdanov:2016tdf}
	\be
	T_{aa}= p-\m^{a}\r^{a}
	=\frac{B^{2}u_{+}^{2}\big(\log(u_{+}\Lambda^{-2})-2\big)+4}{8\kappa^{2}u_{+}^{2}} ,\quad a=1,2,3\ ,
	\label{taa}
	\ee
	where it should be noted that the repeated indices are not summed.
	The temperature, entropy density and magnetic flux density can be computed by standard methods and they are given by
	\be
	T=\frac{8-B^{2}u_{+}^{2}}{8\pi\sqrt{u_{+}}},\quad s=\frac{2\pi}{\kappa^{2}u_{+}^{3/2}},\quad \r^{a}=\frac{B}{4\kappa^2},\quad a=1,2,3\ .
	\label{tsr}
	\ee
	It is straightforward to verify that the quantities above satisfy the first law
	\be
	d\varepsilon= Tds+\sum_{i=1}^{3}\m^{a}d\r^{a}\ .
	\ee
	in which the variable $\m^a$ conjugate to magnetic flux density is given by
	\be
	\m^{a}=B\log(u_{+}\Lambda^{-2}),\quad a=1,2,3\ .
	\label{ma}
	\ee
	Combining \eqref{taa}, \eqref{tsr} and \eqref{ma}, we obtain the pressure as
	\be
	p=-\frac{B^{2}u_{+}^{2}\big(2-3\log(u_{+}\Lambda^{-2})\big)-4}{8\kappa^{2}u_{+}^{2}}\ ,
	\ee
	using which one can show the  Gibbs-Duhem relation $\varepsilon+p=Ts+\sum_{a=1}^{3}\m^{a}\r^{a} $ is satisfied. In the result above, terms proportional to $B^2$ are contributed by magnetic fields. The opposite signs show that there is an interesting competition between them.
	The holographic energy-momentum tensor and the 2-form currents of the electric brane solution can thus be arranged into the standard form of magnetohydrodynamics \footnote{Here we adopt the same convention for the energy-momentum tensor as in \cite{Grozdanov:2016tdf}, which is different from the one used in \cite{Hernandez:2017mch, Meiring:2023wwi}.}
	\bea
	T^{\mu\nu}&=&(\varepsilon+p)u^{\mu}u^{\nu}+p\eta^{\mu\nu}-\sum_{a=1}^{3}\mu^{a}\rho^{a}h^{\mu}_a h^{\nu}_a\ ,
	\nn\\
	J^{a\m\n}&=&2\r^a u^{[\n}h_a^{\m]}\ ,\quad u^\m=\d^\m_0,\quad h^\m_a=\d^\m_a\ .
	\label{tj0}
	\eea
	Note that due to the logarithmic terms present in the thermodynamic quantities, the scaling symmetry \eqref{ss} is broken explicitly. Also the fluid on the boundary of AdS is not conformal as trace of the holographic energy-momentum tensor is nonvanishing
	\be
	-T_{tt}+\sum_{a=1}^3 T_{aa}=-\frac{3B^2}{4\k^2}\ .
	\ee
	To conclude this section, we would like to point out that the thermodynamic quantities obtained above using holographic methods agree with those computed using the improved Wald formalism \cite{whp, Ma:2022nwq, Lu:2013ura}.

	\section{First order hydrodynamics and Kubo formulae}\label{section3}
	Before moving on to the explicit computation of transport coefficients, we first define them in the first order magnetohydrodynamics. In the gravity side, there are three 2-form gauge fields which are dual to three conserved 2-form currents. As noted below \eqref{vs}, a combination of their leading and subleading Fefferman-Graham (FG) expansion coefficients near the AdS boundary denoted as $b^a_{\m\n}$, plays the role of the source for the conserved 2-form current. Invariance of the fluid effective action under the boundary diffeomorphism and 2-form gauge transformation leads to the conservation equations below
	\be
	\nabla_{\m}T^{\m\n}=\sum_{a=1}^3 H^{a\n}{}_{\r\s}J^{a \r\s},\quad \nabla_{\m} J^{a \m\n}=0\ ,
	\ee
	where $H^a=db^a$ \footnote{In this section, $H^a_{\m\n\r}$ denotes the field strength of $b^a_{\m\n}$, which should not be confused with the field strength of $B^a_{MN}$ in the gravity action. }. In magnetohydrodynamics, these two equations are solved order by order in terms of derivatives of fluid variables such as $u^\m, h^\m_a$ and so on. Up to the first order, the energy-momentum tensor and 2-form currents are given by
	\be
	T^{\mu\nu} = T_{(0)}^{\mu\nu}+T_{(1)}^{\mu\nu},\quad J^{a\m\n}=J^{a\m\n}_{(0)}+J^{a\m\n}_{(1)}\ ,
	\ee
	where the zeroth order quantities $T_{(0)}^{\mu\nu},\, J^{a\m\n}_{(0)}$ are given in \eqref{tj0}. Below we derive the expressions for $T_{(1)}^{\mu\nu},\, J^{a\m\n}_{(1)}$. Similar to the previous work \cite{Grozdanov:2016tdf, Hernandez:2017mch}, we consider divergence of the entropy current defined below
	\be
	T S^\m=p u^\m-T^{\m\n}u_{\n}-\m^a J^{\m\n}_a h_{\n}\ .
	\ee
	Using the conservation equations, we find that up to the current order, the divergence of the entropy current is given by
	\be
	\nabla_{\mu}S^{\mu}=-\left[T_{(1)}^{\mu\nu}\nabla_{\mu}(\frac{u_{\nu}}{T})+\sum_{a=1}^3J^{\m\n}_{a(1)}\Big(\nabla_{[\m}(\frac{h^a_{\n]}\mu^a}{T})+\frac{u_{\s}H^{a\s}{}_{\m\n}}{T}\Big)\right]\ ,
	\label{ds}
	\ee
	which ought to be non-negative. We choose to work in the Landau frame where
	$T^{(1)\m\n}u_\n=0$. Thus $\nabla_{\mu}S^{\mu}\ge 0$ allows the general parity-preserving structure of the first order energy-momentum tensor \footnote{In the parity-even case, our $T^{\m\n}_{(1)}$ is more general than the one proposed in \cite{Hernandez:2017mch}.}
	\be
	T_{(1)}^{\mu\nu} = -\sum_{a,\,b=1}^{3}\zeta_{ab}P^{\m\n}_a P^{\a\b}_b\nabla_\a u_\b-2\sum_{\substack{a,\,b=1\\
			a\neq b
		}
	}^{3}\eta_{(ab)} P^{\m\a}_a P^{\n\b}_b\nabla_{(\a} u_{\b)} \ ,
	\label{t1}
	\ee
	where we have defined the projection operator $P^{\m\n}_a:=h_a^\m h_a^\n$.
	
	To derive the first order correction to the 2-form current, we work in a gauge such that $J^{\m\n}_{a (1)} u_{\n}=0$ \cite{Grozdanov:2016tdf}. Therefore, the first order 2-form current  can be expressed as $J^{\m\n}_{a (1)}=h^{\m}_b h^{\n}_c J^{~bc}_{a(1)}$, in terms of which the second term in \eqref{ds} becomes
	\be
	\sum_{a=1}^3 J_{a(1)}^{~bc} S^a{}_{bc},\quad  S^a{}_{bc}= h^\m_b h^\n_c\left(\nabla_{[\m}(\frac{\m^a h^a_{\n]}}{T})+\frac{u_\s H^{a\s}{}_{\m\n}}{T}\right)\ .
	\ee
	It turns out that it is more convenient to define
	\be
	\widetilde{J}_{ab(1)}=\frac12 \epsilon_{a cd} J^{~cd}_{b(1)},\quad \widetilde{S}^{ab}=\frac12 \epsilon^{acd} S^b{}_{cd}\ .
	\ee
	We can then parameterize $\widetilde{J}_{ab(1)}$ as
	\be
	\widetilde{J}_{ab(1)}= -r_{(ab)} T \widetilde{S}_{(ab)}-r_{[ab]} T \widetilde{S}_{[ab]}\ ,
	\label{j1}
	\ee
	where the repeated indices are not summed. Using \eqref{t1} and \eqref{j1}, the semi-positivity of entropy production translates into the semi-positivity of transport coefficients as
	\be
	\eta_{(ab)}\ge 0,\quad r_a\ge 0,\quad r_{(ab)}\ge 0,\quad r_{[ab]}\ge 0,\quad \boldsymbol{\zeta}_1\ge0,\quad \boldsymbol{\zeta}_2\ge0
	,\quad \boldsymbol{\zeta}_3\ge 0\ ,
	\ee
	where $\boldsymbol{\zeta}_i$ corresponds to principal minors of the bulk viscosity matrix given below
	\be
	\boldsymbol{\zeta}_{1}=\zeta_{11},\ \boldsymbol{\zeta}_{2}=\left(\begin{array}{cc}
		\zeta_{11} & \zeta_{12}\\
		\zeta_{21} & \zeta_{22}
	\end{array}\right),\ \boldsymbol{\zeta}_{3}=\left(\begin{array}{ccc}
		\zeta_{11} & \zeta_{12} & \zeta_{13}\\
		\zeta_{21} & \zeta_{22} & \zeta_{23}\\
		\zeta_{31} & \zeta_{32} & \zeta_{33}
	\end{array}\right)\ .
	\ee
	As we will see, the transport coefficients computed in the next section do satisfy these conditions.
	
	Given the expressions of the first order energy-momentum tensor and 2-form currents, one can relate the
	transport coefficients to the retarded Green's function in the linear response theory. In order to do so, we perturb the metric slightly away from the flat space and transform it to the frequency domain
	\begin{equation}
		g_{\mu\nu}\rightarrow \eta_{\mu\nu}+\delta h_{\m\n}(t),\quad \delta h_{\m\n}(t)=\int\frac{d\omega}{2\pi}e^{-i\omega t}\delta h_{\mu\nu}(\omega)\ .
	\end{equation}
	Below for the purpose of computing the first order transport coefficients, it suffices to consider perturbations depending only on time.
	From \eqref{t1}, we obtain the perturbed energy-momentum tensor in the frequency domain as
	\bea
	{\rm Im}\delta T_{aa}(\o) &=&\frac{\o}{2}\sum_{b=1}^3 \zeta_{ab}\delta h_{bb}(0)+{\cal O}(\o^2, \delta h^2),\quad  a=1,2,3\ ,
	\nn\\
	{\rm Im}\delta T_{ab}(\o) &=&  \omega\,\eta_{(ab)}\delta h_{ab}(0)+{\cal O}(\o^2, \delta h^2),\quad  a\ne b;\ a,b=1,2,3\ .
	\label{dt}
	\eea
	On the other hand, we
	define the dual of the 2-form source
	\be
	\widetilde{b}^{ab}=\frac12\epsilon^{acd}b^b_{cd }\ ,
	\ee
	and consider it to be a small perturbation
	\be
	\widetilde{b}_{ab}=\d\widetilde{b}_{ab}(t),\quad \d\widetilde{b}_{ab}(t)=\int\frac{d\omega}{2\pi}e^{-i\omega t}\delta\widetilde{b}_{ab}(\omega)\ .
	\label{db}
	\ee
	From \eqref{j1}, we obtain the induced perturbation of the 2-form currents in the frequency domain as
	\be
	{\rm Im} \d\widetilde{J}_{ab}(\o) =\o r_{(ab)} \delta \widetilde{b}_{(ab)}(0)+ \o r_{[ab]}\delta \widetilde{b}_{[ab]}(0)+{\cal O}(\o^2,\d\widetilde{b}^2)\ ,
	\label{ImdJ}
	\ee
	where $a,b$ range from 1 to 3 and are not summed. Again we have omitted terms that are irrelevant for the derivation of Kubo formulae. We recall that in the linear response theory,
	the retarded 2-point Green’s functions are defined as
	\begin{align}
		\delta T_{\mu\nu}(\omega,\vec{k}) & =-\frac{1}{2}G_{\mu\nu}^{TT,\lambda\sigma}(\omega,\vec{k})\delta h_{\lambda\sigma}(\omega,\vec{k})+\mathcal{O}(\delta h^{2},\delta \widetilde{b}^{2}),\nonumber \\
		\delta\tilde{J}_{ab}(\omega,\vec{k}) & =-\frac{1}{2}\sum_{c,d}G_{ab}^{\tilde{J}\tilde{J},(cd)}(\omega,\vec{k})\delta\tilde{b}_{(cd)}(\omega)-\frac{1}{2}\sum_{c,d}G_{ab}^{\tilde{J}\tilde{J},[cd]}(\omega,\vec{k})\delta\tilde{b}_{[cd]}(\omega,\vec{k})+\mathcal{O}(\delta h^{2},\delta \widetilde{b}^{2}).
	\end{align}
	Comparing (28) and (29), we arrive at the Kubo formulae \cite{Hernandez:2017mch}
	\bea
	\zeta_{ab}&=&\lim_{\omega\rightarrow0}\frac{\text{Im}G_{aa}^{TT,bb}(\omega,0)}{-\omega},
	\quad  \eta_{(ab)}=\lim_{\omega\rightarrow0}\frac{\text{Im}G_{(ab)}^{TT,(ab)}(\omega,0)}{-\omega}\ ,\quad (a\neq b)
	\nonumber \\
	r_{(aa)}&=&\lim_{\omega\rightarrow0}\frac{\text{Im}G_{aa}^{\tilde{J}\tilde{J},aa}(\omega,0)}{-2\omega}\ ,\quad
	r_{(ab)}=\lim_{\omega\rightarrow0}\frac{\text{Im}G_{ab}^{\tilde{J}\tilde{J},(ab)}(\omega,0)}{-\omega}\ ,\quad (a\ne b)\ ,\nn\\
	\ r_{[ab]}&=&\lim_{\omega\rightarrow0}\frac{\text{Im}G_{ab}^{\tilde{J}\tilde{J},[ab]}(\omega,0)}{-\omega}\ .
	\label{KuboFormula}
	\eea
	Although in the definition of the bulk viscosity, $\zeta_{ab}$ is not required to be symmetric with respect to $a,b$. The Onsager relation implied by the Kubo formula dictates that $\zeta_{ab}=\zeta_{(ab)}$. Thus totally, there are 21 transport coefficients, including 9
	resistivity coefficients $r_{(ab)}$ and $r_{[ab]}$,
	6 shear viscosity coefficients $\eta_{(ab)}$ and 6 bulk viscosity coefficients $\zeta_{(ab)}$.
	The conformality requires vanishing of the trace of $T^{\m\n}_{(1)}$, which means $\sum_{a=1}^3\zeta_{ab}=0$ for any $b$. As we shall see, this relation is indeed satisfied in our holographic model.

	\section{Transport coefficients} \label{section4}
	In this section, we will compute all the transport coefficients in the first order magnetohydrodynamics using holography. For this purpose, we consider perturbations of the fields
	\begin{equation}
		G_{MN}\rightarrow \bar{G}_{MN}+\delta G_{M N}(t,u),\quad B_{MN}^{a}\rightarrow \bar{B}_{MN}^{a}+\delta B_{MN}^{a}(t,u)\ ,
	\end{equation}
	where $\bar{G}_{MN}$ and $\bar{B}_{MN}^{a}$ comprise the background solution \eqref{mb2}.
	To simplify the computation, we impose the  radial gauge $\delta G_{u M}=\delta B^a_{u M}=0$
	as in \cite{Grozdanov:2017kyl}. We will then solve the linearized Einstein equation and the 2-form field equations with appropriate boundary conditions. Upon substituting the perturbed solutions into the holographic energy-momentum tensor and 2-form currents \eqref{HTHJ2},  we can read off the transport coefficients by comparing to \eqref{dt} and \eqref{ImdJ}.

	We find that the fluctuations can be arranged into four decoupled channels
	\begin{align}
		r_{(ab)}: & \ \delta\widetilde{B}_{(ab)}, &\ a,b=1,2,3\ ,
		\nn \\
		\ r_{[ab]}: & \ \delta G_{ t a},\,\delta\widetilde{B}_{[ab]}, &\ a\ne b\ ;\ a,b=1,2,3\ ,
		\nn \\
		\eta_{(ab)}: & \ \delta G_{a b},\,\delta B_{tb}^{a}, & a\ne b\ ;\ a,b=1,2,3\ ,
		\nn \\
		\zeta_{ab}: & \ \delta G_{t t},\,\delta G_{a a},\,\delta B_{ta}^{a},  & \ a=1,2,3\ ,
		\label{Channel}
	\end{align}
	where we also exhibit the relevant transport coefficients. Note that the repeated indices are not summed and we have dualized the spatial components of $B^a_{bc}$ to $\wt{B}^{ab}=\ft12\epsilon^{acd}B^b_{cd}$ and lowered the indices using $\delta_{ab}$, so that $\wt{B}_{ab}:=\wt{B}^{ab}$. We find that within each channel, the perturbations are governed by just one dynamical second-order differential equation.

	\subsection{Resistivity coefficients $r_{(ab)}$ and $r_{[ab]}$}
	For simplicity, we employ the dimensionless coordinate $z =u/u_+$. For the $r_{(ab)}$- and $r_{[ab]}$-channels, after plugging the perturbations of fields, we obtain the spatial components of the Hodge dual of the perturbed holographic 2-form currents
	\be
	\d\widetilde{J}_{ab}=\lim_{z\rightarrow 0}\frac{z}{2\kappa^{2}}\delta\widetilde{B}_{ab}'(z)\ ,
	\label{djdb}
	\ee
	which means that we need the behavior of $\delta\widetilde{B}_{ab}$ near
	the AdS boundary $z\rightarrow0$ .  Below we give a detailed derivation of
	the resistivity $r_{(ab)}$, for $a, b=1,2,3$.
	To compute $r_{(ab)}$, we need to solve for $\delta\widetilde{B}_{(ab)}(t, z)$.
	Switching to the frequency domain
	\be
	\d\wt{B}_{(ab)}(t,z)=\int\frac{d\omega}{2\pi}e^{-i\omega t}\d\wt{B}_{(ab)}(\omega,z)\ ,
	\ee
	and after a careful study of the linearized 2-form field equations, we find that $\d\wt{B}_{(ab)}(\omega,z)$ satisfies \footnote{We have checked the  equation below by comparing to \cite{Grozdanov:2017kyl}. In fact, after setting ${\cal V}={\cal W}=-\ft12{\rm ln}u$ and $F\rightarrow F/u$,  the Eq.(B.9) of \cite{Grozdanov:2017kyl} becomes the same as our equation \eqref{Baa}. }
	\be
	\delta\widetilde{B}_{(ab)}''+(\frac{1}{z}+\frac{h'(z)}{h(z)})\delta\widetilde{B}_{(ab)}'+\frac{\omega^{2}u_+}{4zh^{2}}\delta\widetilde{B}_{(ab)}=0\ ,
	\label{Baa}
	\ee
	where $h(z):=F(u)=1-z^{2}+\frac{z^{2}u_{B}^{2}}{4}\log z$ is the blackening factor in the background solution \eqref{mb2} and "prime" denotes the derivative with respect to $z$.  Since the background black brane solution is invariant under the rescaling
	of parameters such as $u_{+}\rightarrow u_{+}/\lambda^{2},\ B\rightarrow B\lambda^{2}$, physically inequivalent quantities should be characterized by the scaling invariant variable $u_{B}=u_{+}B$, in terms of which, the scaling invariant temperature is given by
	\be
	T_B:=\frac{T}{\sqrt{B}}=\frac{8-u_{B}^{2}}{8\pi\sqrt{u_{B}}},\quad 0<u_{B}<2\sqrt{2}\ .
	\label{tb}
	\ee
	We apply the Wronskian method  \cite{Davison:2015taa} to
	solve for $\delta\widetilde{B}_{(ab)}$. Firstly, we consider the time-independent solution by setting $\o=0$ in \eqref{Baa}. The resulting equation admits two solutions, of which the one denoted by $\bb_{(ab)}^{(-)}$ is regular on the horizon and tends to a constant at infinity. Using the Wronskian method, the second solution $\bb_{(ab)}^{(+)}$ can be expressed as
	\begin{equation}
		\boldsymbol{b}_{(ab)}^{(+)}(z)=\boldsymbol{b}_{(ab)}^{(-)}(z)\int_{z_0}^z \frac{dy} {\big(\boldsymbol{b}_{(ab)}^{(-)}(y)\big)^{2} y h(y)}\ ,
	\end{equation}
	where $z_0$ is an arbitrary number between 0 and $1$. Different choice of $z_0$ corresponds to shifting $\boldsymbol{b}_{(ab)}^{(+)}(z)$ by a $\text{constant} \times \boldsymbol{b}_{(ab)}^{(-)}(z)$. According to our assumption, the near boundary expansion of $\boldsymbol{b}_{(ab)}^{(-)}(z)$ does not contain a ${\rm ln}z$ term, thus different choice of $z_0$ will not affect the value of resistivity dictated by \eqref{djdb}. When there exists no $\boldsymbol{b}_{(ab)}^{(-)}(z)$ that is regular everywhere, the choice of $z_0$ becomes important.
	
	Near the boundary and the horizon,  $\boldsymbol{b}^{(+)}_{(ab)}$ behaves as
	\begin{equation}
		\boldsymbol{b}_{(ab)}^{(+)}(z)=\begin{cases}
			[\boldsymbol{b}_{(ab)}^{(-)}(0)]^{-1}\text{ln}z+{\rm finite} & \text{for }z\sim0\ ,\\
			-[2\pi T\sqrt{u_{+}}\boldsymbol{b}_{(ab)}^{(-)}(1)]^{-1}\text{ln}(1-z)+{\rm finite} & \text{for }z\sim 1\ ,
		\end{cases}
		\label{baahb}
	\end{equation}
	where $T$ is the temperature of the magnetic brane solution.
	Up to first order in the small frequency $\o\ll 1$, the solution of \eqref{Baa} in the frequency domain
	can be written as \cite{Davison:2015taa,Grozdanov:2017kyl}
	\begin{equation}
		\delta\widetilde{B}_{(ab)}(\omega,z)=\boldsymbol{b}_{(ab)}^{(-)}(z)+\alpha(\omega)\boldsymbol{b}_{(ab)}^{(+)}(z)+\mathcal{O}(\omega^{2})\ .
		\label{baasol}
	\end{equation}
	On the other hand, we require the solution to \eqref{Baa} to satisfy the infalling boundary condition near the horizon \cite{Son:2002sd,Herzog:2002pc,Hartnoll:2009sz}. This means in the near the horizon expansion, the leading order solution takes the form
	\begin{equation}
		\d\wt{B}_{(ab)}(\omega,z)\approx\bb_{(ab)}^{(-)}(1)(1-z)^{-\frac{i\omega}{4\pi T}}=\boldsymbol{b}_{(ab)}^{(-)}(1)\Big(1-\frac{i\omega}{4\pi T}\text{ln}(1-z)\Big)+\mathcal{O}(\omega^{2})\ .
		\label{baain}
	\end{equation}
	Matching \eqref{baasol} with \eqref{baain}, we determine the coefficient
	$\alpha(\omega)$ to be
	\be
	\alpha(\o)=\frac{i\omega}{2}\sqrt{u_{+}}[\boldsymbol{b}_{(ab)}^{(-)}(1)]^{2}\ ,
	\ee
	which when plugged back into \eqref{baasol}, implies that near the AdS boundary
	\be
	\delta\widetilde{B}_{(ab)}(z)=\boldsymbol{b}_{(ab)}^{(-)}(0)\Big(1+\frac{i\omega}{2}\sqrt{u_{+}}\text{ln}z\ \mathcal{B}_{(ab)}\Big)+\mathcal{O}(\omega^{2})\ .
	\label{dBaa}
	\ee
	In the expression above, we introduced the notation
	\be
	\mathcal{B}_{(ab)}=\Big[\frac{\bb_{(ab)}^{(-)}(1)}{\boldsymbol{b}_{(ab)}^{(-)}(0)}\Big]^{2}.
	\ee
	Utilizing \eqref{dBaa} and the fact that the source of the 2-form current is a combination of the first two Fefferman-Graham coefficients of the bulk 2-form fields, i.e. $\delta\tilde{b}_{(ab)}=\delta\tilde{B}_{(ab)}^{(0)}+2\text{ln}\Lambda\ \delta\tilde{B}_{(ab)}^{(1)}$,
	we find that in the frequency domain
	\be
	\boldsymbol{b}_{(ab)}^{(-)}(0) =\delta\widetilde{b}_{(ab)}+\mathcal{O}(\omega)\ .
	\ee
	The perturbed holographic 2-form current is given by
	\begin{equation}
		\d\wt{J}_{(ab)}=\lim_{z\rightarrow 0}\frac{z}{2\kappa^{2}}\delta\wt{B}_{(ab)}'=\frac{i\omega\sqrt{u_{+}}}{4\kappa^{2}}\mathcal{B}_{(ab)}\delta\wt{b}_{(ab)}+\mathcal{O}(\omega^{2})\ ,
	\end{equation}
	from which we read off the resistivity $r_{(ab)}$ according to \eqref{ImdJ}
	\begin{equation}
		r_{(ab)}=\frac{\sqrt{u_{+}}}{4\kappa^{2}}\mathcal{B}_{(ab)}\ .
	\end{equation}
	Now the evaluation of $r_{(ab)}$ boils down to compute the ratio between $\bb^{(-)}_{(ab)}(1)$ and $\bb^{(-)}_{(ab)}(0)$. Since $\bb^{(-)}_{(ab)}(z)$ satisfies \eqref{Baa}
	with $\o=0$, it is solved directly yielding
	\begin{equation}
		\bb_{(ab)}^{(-)}(z)=c_{1}+c_{2}\int_{z_0}^z\frac{dy}{yh(y)}\ ,
	\end{equation}
	where $c_{1}$ and $c_{2}$ are integration constants. The near-boundary
	and the near-horizon expansions of the second term are given by
	\begin{equation}
		\int_{z_0}^z\frac{dy}{yh(y)}=\begin{cases}
			{\rm ln}z+\cdots \  & \text{for}\ z\sim 0\ ,
			\\
			\frac{4{\rm ln}\left(1-z\right)}{B^{2}u_{+}^{2}-8}+\cdots & \text{for}\ z\sim 1\ .
		\end{cases}
	\end{equation}
	Since we demand $\bb_{(ab)}^{(-)}(z)$ to be regular everywhere,
	it means we must set $c_{2}=0$. We thus obtain $\bb_{(ab)}^{(-)}(z)=c_{1}$ and
	\begin{equation}
		\mathcal{B}_{(ab)}=\Big[\frac{\boldsymbol{b}_{(ab)}^{(-)}(1)}{\boldsymbol{b}_{(ab)}^{(-)}(0)}\Big]^{2}=\frac{c_{1}}{c_{1}}=1\quad\Rightarrow \quad r_{(ab)}=\frac{\sqrt{u_{+}}}{4\kappa^{2}}\ .
	\end{equation}

\begin{figure}[!ht]
		\centering
		\includegraphics[scale=1.0]{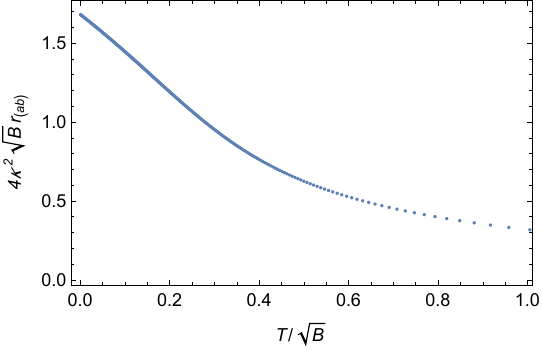}
		\caption{The profile of the scaling invariant resistivity $4\k^2\sqrt{B}r_{(ab)}$.
		}\label{rab1}
	\end{figure}

	The scaling invariant resistivities $r_{(ab)}$ are given by
	\be
	4\kappa^{2}\sqrt{B}r_{(ab)}  =\sqrt{u_{B}},\quad a,b=1,2,3\ .
	\ee
	The behavior of scaling invariant resistivity $r_{(ab)}$ is shown in Fig.\ref{rab1} .
	In fact, from \eqref{tb}, the low $T_B$ and high $T_B$ limit of $4\k^2 \sqrt{B} r_{(ab)}$ can be obtained  analytically as follows
	\bea
	T_B &\rightarrow& 0,\quad 4\k^2 \sqrt{B} r_{(ab)}\rightarrow 2^{3/4}-\frac{\pi T_B}{\sqrt{2}}\ ,
	\nn\\
	T_B &\rightarrow&\infty,\quad 4\k^2 \sqrt{B} r_{(ab)}\rightarrow \frac1{\pi T_B}\ .
	\eea

	\subsubsection{$r_{[ab]}$  with $B=0$}
	We now turn to the computation of the antisymmetric part of the resistivity matrix. We find that the situation with $B=0$ is rather different from the case with nonzero $B$. We thus separate these two cases and first analyze the $B=0$ case which turns out to be much simpler.  In terms of the dimensionless radial coordinate $z=u/u_+$, the relevant linearized field equations are given by
	\footnote{We have checked the equation below by comparing to \cite{Grozdanov:2017kyl}. In fact, after setting $B=0$, ${\cal V}={\cal W}=-\ft12{\rm ln}u$ and $F\rightarrow F/u$,  the Eq.(3.50) of \cite{Grozdanov:2017kyl} coincides with our equation \eqref{rabB0}. }
	\be
	\delta\widetilde{B}_{[ab]}''(z)+(\frac{1}{z}+\frac{f'(z)}{f(z)})\delta\widetilde{B}_{[ab]}'(z)+\frac{\omega^{2}u_{+}}{4zf^{2}}\delta\widetilde{B}_{[ab]}(z)=0\ ,
	\label{rabB0}
	\ee
	where $f(z)=1-z^2$ and ``prime" denotes derivative with respect to $z$. Accordingly, the aforementioned time-independent solution $b_{[ab]}^{(\pm)}(z)$ will obey the following equation
	\be
	\boldsymbol{b}_{[ab]}^{(\pm)''}(z)+(\frac{1}{z}+\frac{h'(z)}{h(z)})\boldsymbol{b}_{[ab]}^{(\pm)'}(z)=0\ ,
	\ee
	which can be solved directly, giving rise to
	\be
	\boldsymbol{b}_{[ab]}^{(-)}(z)=c_{1},  \quad \boldsymbol{b}_{[ab]}^{(+)}(z)=c_{1}(\log\frac{z}{\sqrt{1-z^{2}}}+c_{2})\ ,
	\ee
	where $c_2$ is a constant to be determined later.
	Near the AdS boundary and the horizon, $\boldsymbol{b}_{[ab]}^{(+)}$ behaves as
	\be
	\boldsymbol{b}_{[ab]}^{(+)}=\begin{cases}
		c_{1}(\log z+c_2+\frac{z^{2}}{2})+\cdots & \text{for }z\sim0\ ,\\
		c_{1}\big(-\frac{1}{2}\log(1-z)-\frac{1}{2}\log2+c_2-\frac{3}{4}(1-z)\big)+\cdots & \text{for }z\sim1\ .
	\end{cases}
	\ee
	
	Up to the first order in the
	small frequency $\omega\ll1$, the solution of \eqref{rabB0} can be constructed as
	\begin{equation}
		\delta\widetilde{B}_{[ab]}(\omega,z)=\boldsymbol{b}_{[ab]}^{(-)}(z)+\alpha(\omega)\boldsymbol{b}_{[ab]}^{(+)}(z)+\mathcal{O}(\omega^{2})\ .
		\label{B0dB1}
	\end{equation}
	The infalling boundary condition near the horizon implies $\delta\tilde{B}_{[ab]}(\omega,z)$ should take the form
	\bea
	\delta\widetilde{B}_{[ab]}(\omega,z)	&=&\boldsymbol{b}_{[ab]}^{(-)}(1)(1-z)^{-\frac{i\omega}{4\pi T}}\Big(1-i\omega\frac{3}{8}\sqrt{u_{+}}(1-z)+\cdots \Big)\ , \cr
	&=&\boldsymbol{b}_{[ab]}^{(-)}(1)\Big(1-\frac{i\omega}{4\pi T}\text{ln}(1-z)-i\omega\frac{3}{8}\sqrt{u_{+}}(1-z)+\cdots \Big)\ .
	\label{B0dB2}
	\eea
	Matching \eqref{B0dB1} with  \eqref{B0dB2}, we determine the coefficients $c_2$ and $\alpha(\omega)$ to be
	\be
	c_2=\frac{\log2}{2},\quad \alpha=\frac{i\omega}{2\pi T}\ ,
	\ee
	where $T=\frac{1}{\pi \sqrt{u_+}}$.
	With these results in hand, we find that near the boundary of AdS
	\be
	\delta\wt{B}_{[ab]}(\omega,z)\approx c_{1}\Big(1+c_2+\frac{i\omega}{2\pi T}\log z\Big)\ .
	\ee
	The holographic 2-form current is thus obtained as
	\be
	\delta\wt{J}_{[ab]}=\lim_{z\rightarrow0}\frac{z}{2\kappa^{2}}\delta\wt{B}_{[ab]}'=i\omega\frac{c_{1}\sqrt{u_{+}}}{4\kappa^{2}}+\mathcal{O}(\omega^{2})\ .
	\ee
	As the source of the 2-form current is  given by $\delta\wt{b}_{[ab]}=c_1+{\cal O}(\o)$, we obtain the  resistivity coefficients $r_{[ab]}$ with $B=0$ as
	\be
	r_{[ab]}=\frac{\sqrt{u_{+}}}{4\kappa^{2}}=\frac{1}{4\pi \k^2 T}\ ,
	\ee
	which is the same as the resistivity coefficients $r_{(ab)}$ at $B=0$.
	\subsubsection{$r_{[ab]}$  with $B\ne0$}
	We now turn to the computation of the antisymmetric part of the resistivity matrix with $B\ne0$. In terms of the dimensionless coordinate $z$, the relevant linearized field equations are \footnote{We have checked the equation below by comparing to \cite{Grozdanov:2017kyl}. In fact, after setting ${\cal V}={\cal W}=-\ft12{\rm ln}u$ and $F\rightarrow F/u$,  the Eq.(3.50) of \cite{Grozdanov:2017kyl} coincides with our equation \eqref{Bab2}. }
	\be
	\delta\wt{B}_{[ab]}'' +(\frac{1}{z}+\frac{h'}{h})\delta\wt{B}_{[ab]}'+(\frac{\omega^{2}u_+}{4zh^{2}}-\frac{u_B^{2}}{2h})\delta\wt{B}_{[ab]}=0\ .
	\label{Bab2}
	\ee
	where $ h(z)=1-z^{2}+\frac{z^{2}u_{B}^{2}}{4}\log z$ and ``prime" denotes derivative with respect to $z$. Accordingly, the two time-independent solutions $b_{[ab]}^{(\pm)}(z)$ satisfy the following equations
	\be
	\bb^{(\pm)''}_{[ab]}(z) +(\frac{1}{z}+\frac{h'(z)}{h(z)})\bb_{[ab]}^{(\pm)'}(z)-\frac{u_{B}^{2}}{2h(z)}\bb^{(\pm)}_{[ab]}(z)=0\ ,
	\label{Bz2}
	\ee
	which can be solved straightforwardly, yielding the results
	\be
	\boldsymbol{b}_{[ab]}^{(-)}(z)=c(u_{B}^{2}-8+2u_{B}^{2}\log z),\ \ \ \boldsymbol{b}_{[ab]}^{(+)}(z)=\boldsymbol{b}_{[ab]}^{(-)}(z)\Big(\int_{0}^{z}\frac{dy}{yh(y)[\boldsymbol{b}_{[ab]}^{(-)}(y)]^{2}}+\gamma\Big)\ .
	\ee
	Note that we are able to choose the lower limit of the integral to 0 is due to the fact that
	when $B\neq 0$, the integral converges at $z=0$, which is not the case when $B=0$.
	Near the boundary and the horizon, $\boldsymbol{b}_{[ab]}^{(-)}$ behaves as
	\be
	\boldsymbol{b}_{[ab]}^{(-)}(z)=\begin{cases}
		c(u_{B}^{2}-8+2u_{B}^{2}\log z)\ , & \text{for }z\sim0\ ,\\
		c(u_{B}^{2}-8)+2c(z-1)u_{B}^{2}+...\ , & \text{for }z\sim1\ ,
	\end{cases}
	\ee
	which indicates that $\bb_{[ab]}^{(-)}$ is no longer regular everywhere as it diverges logarithmically approaching the AdS boundary. This is why the $B\neq0$ case is drastically different from the $B=0$ case. This subtlety was not noticed in the previous work \cite{Grozdanov:2017kyl} possibly due to the lack of the analytic background solution. In fact, had we naively followed the same analysis as before, numerically we would obtain similar plots for $r_{[ab]}$ as in \cite{Grozdanov:2017kyl}. The reason is that we integrate the differential equation up to a very small value of $z$ but not strictly 0, and it is hard to tell the logarithmic divergence in $\bb^{(-)}_{[ab]}$ numerically.

\begin{figure}[!ht]
		\centering
		\includegraphics[scale=1.0]{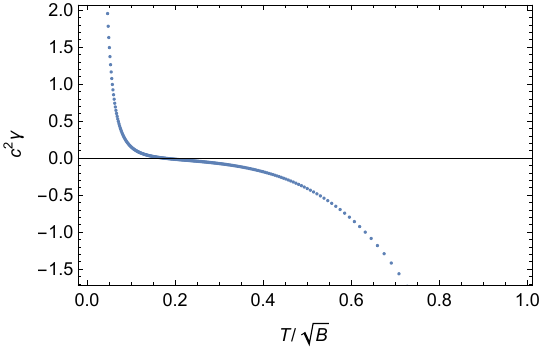}
		\caption{The dependence of $\gamma$ on $T_B$. At the critical value of $u^c_B\approx 1.5896878$ and $T_B=\frac{8-(u_{B}^{c})^{2}}{8\pi\sqrt{u_{B}^{c}}}\approx0.172711 \  $, $\g=0$.  When $T_B$ is small, $c^2\gamma$ approaches positive infinity.  }\label{gamma}
	\end{figure}

\begin{figure}[!ht]
		\centering
		\includegraphics[scale=1.1]{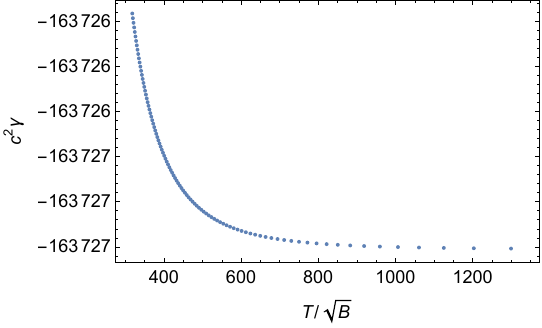}
		\caption{The dependence of $\gamma$ on large $T_B$. When $T_B$ is large,
			$c^2\gamma$ approaches $-163726.65$. }\label{BigTBgamma}
	\end{figure}

	On the other hand,
	$\boldsymbol{b}_{[ab]}^{(+)}$ behaves as
	\be
	\boldsymbol{b}_{[ab]}^{(+)}(z)=\begin{cases}
		\gamma\boldsymbol{b}_{[ab]}^{(-)}(z)-\frac{1}{2cu_{B}^{2}}+\text{higher order terms} & \text{for }z\sim0\ ,\\
		\frac{-\log(1-z)}{2\pi T\boldsymbol{b}_{[ab]}^{(-)}(1)\sqrt{u_{+}}}+c(u_{B}^{2}-8)(\gamma+\delta)+\text{higher order terms}\ , & \text{for }z\sim1\ ,
	\end{cases}
	\ee
	where $\delta$ is the constant appearing in the near horizon expansion
	\be
	\delta=\int_{0}^{1} dz\Big(\frac{1}{zh(z)[\boldsymbol{b}_{[ab]}^{(-)}(z)]^{2}}-\frac{4}{c^{2}(u_{B}^{2}-8)^{3}(z-1)}\Big)\ ,
	\ee
	which is inversely proportional to $c^2$.
	Up to first order in the
	small frequency $\omega\ll1$, the solution of \eqref{Bab2} can still be written as
	\begin{equation}
		\delta\widetilde{B}_{[ab]}(\omega,z)=\boldsymbol{b}_{[ab]}^{(-)}(z)+\alpha(\omega)\boldsymbol{b}_{[ab]}^{(+)}(z)+\mathcal{O}(\omega^{2})\ .
		\label{babsol}
	\end{equation}
	The infalling boundary condition near the horizon implies $\delta\widetilde{B}_{[ab]}(\omega,z)$ should take the form
	\bea
	\delta\widetilde{B}_{[ab]}(\omega,z)	&=&\boldsymbol{b}_{[ab]}^{(-)}(1)(1-z)^{-\frac{i\omega}{4\pi T}}\big(1+b_{1}(1-z)+{\cal O}((1-z)^2)\big)\ , \cr	
	&=&\boldsymbol{b}_{[ab]}^{(-)}(1)\big(1-\frac{i\omega}{4\pi T}\text{ln}(1-z)+\cdots\big)\ ,
	\label{babBC}
	\eea
	where $b_{1}=-\frac{2u_{B}^{2}}{u_{B}^{2}-8}+\frac{i\sqrt{u_{+}}\omega\left(13u_{B}^{2}-24\right)}{\left(u_{B}^{2}-8\right){}^{2}}$. By matching \eqref{babsol} with \eqref{babBC}, we determine the coefficient $\gamma$ and $\alpha(\omega)$  to be
	\be
	\gamma=-\delta\ ,\quad \alpha(\omega)=\frac{i\omega}{2}\sqrt{u_{+}}[\boldsymbol{b}_{[ab]}^{(-)}(1)]^{2}\ .
	\ee
	We  plot $\gamma$ as a function of $T_B$ in Fig. \ref{gamma} and Fig. \ref{BigTBgamma}. One can see that at some critical $T^c_B:=T_B(u^c_B)$, $\g=0$. When $T_B$ is small, $\gamma$ approaches positive infinity at certain rate. Due to the numerical instability in this region, we have not found a reliable fitting. When $T_B$ is large, $c^2\gamma$ approaches a negative constant.
	
\begin{figure}[!ht]
		\centering
		\includegraphics[scale=1.2]{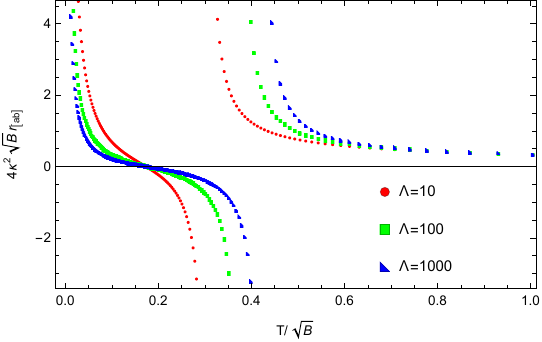}
		\caption{The profile of the scaling invariant resistivity $4\kappa^{2}\sqrt{B}r_{[ab]}$ as a function of the scaling invariant temperature $T_B=T/\sqrt{B}$.
		}\label{rabnew}
	\end{figure}

	Using the results above, we find that near the AdS boundary
	\be
	\delta\widetilde{B}_{[ab]}=\boldsymbol{b}_{[ab]}^{(-)}(z)+\frac{i\omega}{2}\sqrt{u_{+}}[\boldsymbol{b}_{[ab]}^{(-)}(1)]^{2}\left(\gamma\boldsymbol{b}_{[ab]}^{(-)}(z)-\frac{1}{2cu_{B}^{2}}\right)+\text{higher order terms}\ ,
	\label{Babsol2}
	\ee
	from which we obtain the Hodge dual of holographic 2-form current as
	\be
	\delta\wt{J}_{[ab]}	=\lim_{z\rightarrow0}\frac{z}{2\kappa^{2}}\delta\wt{B}_{[ab]}'=\frac{cu_{B}^{2}\left(2+i\omega\sqrt{u_{+}}\gamma[\boldsymbol{b}_{[ab]}^{(-)}(1)]^{2}\right)}{2\kappa^{2}}+\mathcal{O}(\omega^{2})\ .
	\label{J1ab}
	\ee
	The source of the dual 2-form currents is read off from the first two FG expansion coefficients
	of $\widetilde{B}_{[ab]}$
	\be
	\delta\widetilde{b}_{[ab]}=\delta\widetilde{B}_{[ab]}^{(0)}+2\text{ln}\Lambda\ \delta\widetilde{B}_{[ab]}^{(1)}\ ,
	\label{source}
	\ee
	where
	\bea
	\delta\widetilde{B}_{[ab]}^{(0)}&=&c(u_{B}^{2}-8)+\frac{i\omega}{2}\sqrt{u_{+}}[\boldsymbol{b}_{[ab]}^{(-)}(1)]^{2}\big(c\gamma(u_{B}^{2}-8)-\frac{1}{2cu_{B}^{2}}\big)\ ,
	\nn\\
	\delta\widetilde{B}_{[ab]}^{(1)}&=&2cu_{B}^{2}\big(1+\gamma\frac{i\omega}{2}\sqrt{u_{+}}[\boldsymbol{b}_{[ab]}^{(-)}(1)]^{2}\big)\ .
	\eea
	Substituting \eqref{source} into \eqref{ImdJ}, we obtain the antisymmetric part of the resistivity matrix
	\be
	r_{[ab]}  =\frac{c^2 \gamma  \sqrt{u_+}   u_B^2 \left(u_B^2-8\right){}^2}{2 \kappa ^2 \left(u_B^2 (4 \log \Lambda +1)-8\right)}\ ,\quad a,b=1,2,3\ .
	\label{res2}
	\ee
	The behavior of scaling invariant resistivity $4\kappa^{2}\sqrt{B}r_{[ab]}$ is shown in Fig. \ref{rabnew}. It can be observed that the resistivity admits two branches separated by $T^d_B=\frac{2 \sqrt[4]{2} \log \Lambda }{\pi  (4 \log \Lambda +1)^{3/4}}$ corresponding to $u_B=\frac{2}{\sqrt{2 \log \Lambda +\frac{1}{2}}}$ .
	When  $T_B>T^d_B$, the resistivity is positive. It diverges as $T_B$ approaches $T^d_B$ and tends to 0 as $T_B\rightarrow \infty$. When $0<T_B<T^d_B$, the resistivity is not positive definite and it hits 0 at $T^c_B$. In between $T_B^c$ and $T_B^d$, counter-intuitively, the resistivity becomes negative and violates the positivity discussed in section \ref{section3}. For the time being, we do not have a deeper understanding why this happens and would like to postpone it to future investigation. Interestingly, the value of $r_{[ab]}$ depends on the UV cutoff scale $\L$ \footnote{Since the equation of motion governing this channel is closely related to the Eq.(3.50) of \cite{Grozdanov:2017kyl}, qualitatively the phenomenon observed here should also appear in
		the transverse resistivity $r_{\bot}$ in the more realistic model with a single electromagnetic field. The fact that \cite{Grozdanov:2017kyl} has missed it is probably due to the lack of an analytic background solution such that it is difficult to observe the $\log z$ behavior in the ``would be" regular solution of the time-independent equation.  }.  From \eqref{res2}, the low $T_B$ and high $T_B$ limit of $4\k^2 \sqrt{B} r_{[ab]}$ can be obtained  analytically
	\bea
	T_B &\rightarrow& 0,\quad 4\k^2 \sqrt{B}r_{[ab]}\rightarrow \frac{128 \sqrt[4]{2} c^2 \gamma  T_B^2}{\log \Lambda }\  ,
	\nn\\
	T_B &\rightarrow&\infty,\quad 4\k^2 \sqrt{B} r_{[ab]}\rightarrow -\frac{16 c^2 \gamma }{\pi ^5 T_B^5}\ .
	\eea

	\subsection{Shear viscosity $\eta_{(ab)}$}
	In the shear channel, the relevant perturbations are
	\be
	\delta G_{a b},\quad \delta B_{tb}^{a}, \quad a\ne b\ ;\ a,b=1,2,3\ .
	\ee
	In fact, perturbations in each shear channel labeled by different values of $a, b$ obey identical equations of motion owing to the symmetry of the background solution. Therefore, we will just elaborate on the computation of one shear viscosity coefficient, say, $\eta_{(12)}$. The other components of $\eta_{(ab)}$ will share the same value. In the $\eta_{12}$ channel, the relevant  equations of motion for the perturbations are \footnote{We have checked the second equation below by comparing to \cite{Grozdanov:2017kyl}. In fact, after setting ${\cal V}={\cal W}=-\ft12{\rm ln}u$ and $F\rightarrow F/u$,  the Eq.(B.1) of \cite{Grozdanov:2017kyl} becomes the same as our equation \eqref{EOMG12}. }
	\begin{align}
		\delta B_{t2}^{1'}& =\delta B_{t1}^{2'}=\frac{B}{2u}\delta G_{\ 2}^{1}\ ,
		\nonumber \\
		\delta G_{\ 2}^{1''} & +(-\frac{1}{u}+\frac{F'}{F})\delta G_{\ 2}^{1'}+(\frac{\omega^{2}}{4uF^{2}}-\frac{B^{2}}{4F})\delta G_{\ 2}^{1}=0\ ,
		\label{EOMG12}
	\end{align}
	where $\d G^1{}_2=\bar{G}^{11}\d G_{12}$ and all the perturbations have been Fourier transformed into frequency domain. Using \eqref{EOMG12}, we find that the holographic energy-momentum tensor in the frequency domain takes the form
	\begin{equation}
		\delta T_{12}(\o,u)=\frac{1}{\kappa^{2}}\lim_{u\rightarrow0}\big(\frac{\sqrt{F}}{u}\delta G_{\ 2}^{1'}+f_{\eta}(u)\delta G_{\ 2}^{1}\big)+\mathcal{O}(\omega^{2})\ ,
	\end{equation}
	where $f_{\eta}(u)=\frac{1}{8u^{2}F}\Big(24F^{\frac{3}{2}}-F\big(24+B^{2}u^{2}\text{ln}(u\Lambda^{-2})\big)-8u\sqrt{F}F'\Big)$. We again solve for $\d G^1{}_2$ using Wronskian approach.
	The equation \eqref{EOMG12} with $\o=0$ admits two independent solutions, of which the one that is regular everywhere is denoted as $\boldsymbol{g}_{\ 2}^{1(-)}(u)$. Unlike $\bb_{[ab]}^-(u)$, the existence of such a regular solution is guaranteed  by the FG expansion of the metric \eqref{FG}. The Wronskian approach allows us to express the second time-independent solution denoted by $\boldsymbol{g}_{\ 2}^{1(+)}(u)$ as
	\begin{equation}
		\boldsymbol{g}_{\ 2}^{1(+)}(u)=\boldsymbol{g}_{\ 2}^{1(-)}(u)\int_{u_{0}}^{u}\frac{ydy}{\big(\boldsymbol{g}_{\ 2}^{1(-)}(y)\big)^{2}F}\ .
	\end{equation}
	Near the AdS boundary and the horizon, $\boldsymbol{g}_{\ 2}^{1(+)}(u)$ behaves as
	\begin{equation}
		\boldsymbol{g}_{\ 2}^{1(+)}(u)=\begin{cases}
			u^{2}[2\boldsymbol{g}_{\ 2}^{1(-)}(0)]^{-1}+\cdots & \text{for }u\sim0\ ,\\
			-u_{+}^{\frac{3}{2}}[2\pi T\boldsymbol{g}_{\ 2}^{1(-)}(u_{+})]^{-1}\text{ln}(u_{+}-u)+\cdots & \text{for }u\sim u_{+}\ .
		\end{cases}
	\end{equation}
	Up to the first order in the small frequency $\o\ll 1$, the time-dependent solution of \eqref{EOMG12} in the frequency can be written as \cite{Davison:2015taa,Grozdanov:2017kyl}
	\be
	\d G^1{}_2(\o, u)=\boldsymbol{g}_{\ 2}^{1(-)}(u)+\b(\o) \boldsymbol{g}_{\ 2}^{1(+)}(u)+{\cal O}(\o^2)\ .
	\label{dg121}
	\ee
	On the other hand, near the horizon the time-dependent solution should satisfy the infalling boundary condition and thus takes the form
	\be
	\d G^1{}_2(\o, u)\approx \boldsymbol{g}_{\ 2}^{1(-)}(u_+)(u_+-u)^{-\frac{i\o}{4\pi T}}=\boldsymbol{g}_{\ 2}^{1(-)}(u_+)\Big(1-\frac{i\o}{4\pi T}{\rm ln}(u_+-u)\Big)+{\cal O}(\o^2)\ .
	\label{dg122}
	\ee
	Comparing \eqref{dg121} with \eqref{dg122}, we  determine
	\be
	\b(\o)=\frac{i\o  [\boldsymbol{g}_{\ 2}^{1(-)}(u_+)]^2 }{2 u_+^{\frac32}}\ .
	\ee
	Substituting the result back into \eqref{dg121} and expanding around $u=0$, we obtain
	\begin{equation}
		\delta G_{\ 2}^{1}(\o, u)=(1+\frac{i\omega u^{2}}{4}u_{+}^{-\frac{3}{2}}\mathcal{G}_{12})\ \delta h_{12}+\cdots\ ,
		\label{dG12}
	\end{equation}
	where  $\mathcal{G}_{12}=\Big[\frac{\boldsymbol{g}_{12}^{(-)}(u_{+})}{\boldsymbol{g}_{12}^{(-)}(0)}\Big]^{2}$ and $\delta h_{12}$ denotes the leading FG expansion coefficient of the metric component $G_{12}$, which is equal to $\boldsymbol{g}_{\ 2}^{1(-)}(0)$.
	Substituting eq.(\ref{dG12}) into $\delta T_{12}$, we extract the
	imaginary part of $\delta T_{12}$ as
	\begin{align*}
		{\rm Im}\delta T_{12} & =\frac{\omega}{2\kappa^{2}u_{+}^{\frac{3}{2}}}\mathcal{G}_{12}\delta h_{12}+\mathcal{O}(\omega^{2})\ ,
	\end{align*}
	which when compared to \eqref{dt}, leads to
	\begin{equation}
		\eta_{(12)}=\frac{\mathcal{G}_{12}}{2\kappa^{2}u_{+}^{\frac{3}{2}}}=\frac{s}{4\pi}\mathcal{G}_{12}\ ,
		\label{shear}
	\end{equation}
	where $s$ is the entropy density of the magnetic brane given in \eqref{tsr}. Due to the SO(3) symmetry, all
	the shear viscosity coefficients are identical, i.e. $\eta_{(ab)}=\frac{s}{4\pi}\mathcal{G}_{12}$.

\begin{figure}[!ht]
		\centering
		\includegraphics[scale=1.0]{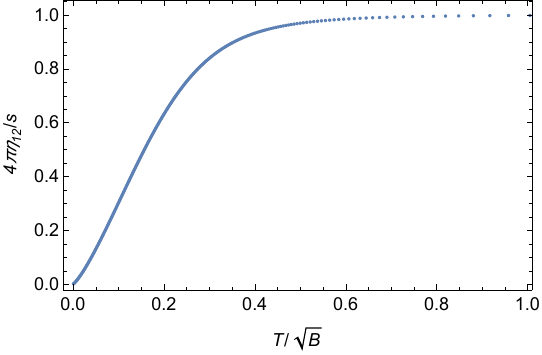}
		\caption{The profile of the scaling invariant viscosity to entropy density ratio $4\pi\eta_{[12]}/s$. The precision of the numerical computation is set by the maximal value $n$ where we truncate the infinite series. This plot is obtained for $n_{\rm max}=10$ which is sufficiently good compared to bigger values of $n_{\rm max}$.
		}\label{etaab}
	\end{figure}

	We now proceed to solve $\boldsymbol{g}_{[12]}^{(-)}$ numerically.
	In terms of the dimensionless coordinate $z=u/u_+$, the equation satisfied by $\boldsymbol{g}_{[12]}^{(-)}$ can be recast into
	\be
	\bg^{(-)''}_{[12]}(z) +(-\frac{1}{z}+\frac{h'(z)}{h(z)})\bg_{[12]}^{(-)'}(z)-\frac{u_{B}^{2}}{4h(z)}\bg^{(-)}_{[12]}(z)=0\ .
	\label{g122}
	\ee
	Regularity of $\bg_{[12]}^{(-)}$ implies that near the horizon of the black brane, it can be expanded as
	$\bg_{[12]}^{(-)}(z)=\sum_{n=0}g_{h(n)}(1-z)^{n}$. Equation \eqref{g122} then determines $g_{h(n)}$ to arbitrary order. Below we list only the first five expansion coefficients
	\begin{align}
		g_{h(1)} & =-\frac{u_{B}^{2}g_{h(0)}}{u_{B}^{2}-8},\ \  \ g_{h,2}=-\frac{u_{B}^{4}g_{h(0)}}{4\left(u_{B}^{2}-8\right)^{2}},\ \ \ g_{h(3)}=-\frac{u_{B}^{2}\left(u_{B}^{4}+6u_{B}^{2}-32\right)g_{h(0)}}{9\left(u_{B}^{2}-8\right)^{3}}\ ,\nonumber \\
		g_{h,4} & =-\frac{u_{B}^{2}\left(35u_{B}^{6}+672u_{B}^{4}-5632u_{B}^{2}+12288\right)g_{h(0)}}{576\left(u_{B}^{2}-8\right)^{4}}\ ,\nonumber \\
		g_{h(5)} & =-\frac{u_{B}^{2}\left(531u_{B}^{8}+22096u_{B}^{6}-244224u_{B}^{4}+999424u_{B}^{2}-1572864\right)g_{h(0)}}{14400\left(u_{B}^{2}-8\right)^{5}}\ .
		\label{gzser}
	\end{align}
	Note that $\bg^{(-)}_{[12]}$ is characterized by just one  free parameter instead of two. The reason is that
	the other free parameter comes with ${\rm ln}(1-z)$ which is divergent on the horizon. In solving \eqref{g122} numerically, we integrate from the horizon to the AdS boundary and then read off the ratio between $\bg^{(-)}_{[12]}(1)$ and $\bg^{(-)}_{[12]}(0)$. Since \eqref{g122} is a homogeneous equation, different choice for  $\bg^{(-)}_{[12]}(1)=g_{h(0)}$ will lead to the same ratio. We thus choose  $g_{h(0)}=1$ for simplicity. In practice, we cannot place the initial value exactly at $z=1$  where the equation \eqref{g122} becomes singular. Instead, the initial value is assigned at $z=1-10^{-9}$. To read off the value of $\bg^{(-)}_{[12]}(z)$ and its first derivative at $z=1-10^{-9}$, we utilize the series expansion truncated at 10th order in $(1-z)$.  Once we acquire the ratio
	$\bg^{(-)}_{[12]}(1)/\bg^{(-)}_{[12]}(0)$, the shear viscosity $\eta_{[12]}$ is obtained via \eqref{shear}. In Fig. \ref{etaab}, we plot the scaling invariant viscosity to entropy density ratio as a function of the scaling invariant temperature $T_B$.
	We observe that when $T_B$ is large, the ratio approaches 1 as expected while when $T_B$ is small, it
	is approximately given by
	\be
	\frac{4\pi}{s}\eta_{12} \sim0.016T_{B}^{2}+0.12T_{B}^{3}+0.63T_{B}^{4}+\cdots\ .
	\ee
	The leading $T^2_B$ behavior in the small $T_B$ region was also reported in \cite{Grozdanov:2017kyl}.

	\subsection{Bulk viscosity $\zeta_{ab}$}
	In this channel, the relevant perturbations are
	\be
	\delta G_{t t},\quad \delta G_{a a},\quad \delta B_{ta}^{a},  \quad \ a=1,2,3\ ,
	\ee
	which are entangled with each other in the linearized field equations.
	To diagonalize the perturbations, we redefine the variables as
	\bea
	\delta G_{1}&=&\bar{G}^{11}\delta G_{11}-\bar{G}^{22}\delta G_{22},\quad \delta G_{2}=\bar{G}^{22}\delta G_{22}-\bar{G}^{33}\delta G_{33}\ ,
	\nn\\
	\delta B_{1}&=&\delta B_{t1}^{1}-\delta B_{t2}^{2},\quad \delta B_{2}=\delta B_{t2}^{2}-\delta B_{t3}^{3}\ ,
	\nn\\
	\delta G_{+}&=&\bar{G}^{11}\delta G_{11}+\bar{G}^{22}\delta G_{22}+\bar{G}^{33}\delta G_{33}\ ,
	\quad \delta B_{+}=\delta B_{t1}^{1}+\delta B_{t2}^{2}+\delta B_{t3}^{3}\ .
	\eea
	In terms of the new variables, the relevant linearized equations become
	\begin{align}
		&\delta  B_{1}'=\frac{B}{2u}\delta G_{1}\ ,\quad \delta B_{2}'=\frac{B}{2u}\delta G_{2}\ , \quad \delta B_{+}'=\frac{B}{4u}(3\delta G_{\ t}^{t}-\delta G_{+})\ ,
		\nn\\
		& \delta G_{1}''+(-\frac{1}{u}+\frac{F'}{F})\delta G_{1}'+(\frac{\omega^{2}}{4uF^{2}}-\frac{B^{2}}{4F})\delta G_{1}=0\ ,
		\nn \\
		&\delta G_{2}''+(-\frac{1}{u}+\frac{F'}{F})\delta G_{2}'+(\frac{\omega^{2}}{4uF^{2}}-\frac{B^{2}}{4F})\delta G_{2}=0\ ,
		\nn \\
		&\delta G_{+}'-\frac{F'}{2F}\delta G_{+}=0,\quad \delta G_{\ t}^{t'}=\frac{\delta G_{+}}{6F^{2}}\left(\omega^{2}+F'(uF'-3F)-B^{2}uF\right)\ .
		\label{EOMbulk}
	\end{align}
	where $\delta G_{\ t}^{t}=\bar{G}^{tt}\delta G_{tt}$. From \eqref{EOMbulk}, we immediately solve for $\delta G_{+}=c_{+}\sqrt{F(u)}$ with $c_{+}$ being an integration constant.
	Substituting $\delta G_{+}$ into the holographic
	energy-momentum tensor and utilizing equations in \eqref{EOMbulk}, we obtain
	\begin{align}
		& \d T_{G_1}:=\delta T_{11}-\delta T_{22}=\frac{1}{\kappa^{2}}\lim_{u\rightarrow0}(\frac{\sqrt{F}}{u}\delta G_{1}'+f_{\eta}\delta G_{1})+\mathcal{O}(\omega^{2})\ ,
		\nonumber \\
		& \d T_{G_2}:=\delta T_{22}-\delta T_{33}=\frac{1}{\kappa^{2}}\lim_{u\rightarrow0}(\frac{\sqrt{F}}{u}\delta G_{2}'+f_{\eta}\delta G_{2})+\mathcal{O}(\omega^{2})\ ,\nonumber \\
		& \d T_{G_+}:=\sum_{a=1}^{3}\delta T_{aa}=\frac{c_{+}}{\kappa^{2}}\lim_{u\rightarrow0}\Big(F^{\frac{1}{2}}f_{\eta}+\frac{B^{2}}{2}+\frac{F'}{2u}-\frac{F'^{2}}{2F}\Big)+\mathcal{O}(\omega^{2})\ ,\label{BulkHT}
	\end{align}
	where $f_{\eta}(u)=\frac{1}{8u^{2}F}\big(-2u\omega^{2}+3F\big(8\sqrt{F}-8+B^{2}u^{2}\text{ln}(u\Lambda^{-2})\big)-8u\sqrt{F}F'\big)$.
	Note that up to the first order in the small frequency $\o$, the imaginary part of $\sum_{a=1}^{3}\delta T_{aa}$ vanishes as $c_{+},f_{\eta},F$ are real. It is noteworthy that $\d G_1$ and $\d G_2$ satisfy the same equations of motion as $\d G^1{}_2$ in the shear channel. Therefore, we expect their solutions should take similar forms as $\d G^1{}_2$. In particular, near the AdS boundary, they should behave as
	\begin{align}
		& \delta G_{\s}=\boldsymbol{g}_{\s}^{(-)}(0)(1+\frac{i\omega u^{2}}{4}u_{+}^{-\frac{3}{2}}\mathcal{G}_{\s})+\mathcal{O}(\omega^{2})\ ,\quad \s=1,2
		\label{Gi}
	\end{align}
	where $\mathcal{G}_{\s}=\Big[\frac{\boldsymbol{g}_{\s}^{(-)}(u_{+})}{\boldsymbol{g}_{\s}^{(-)}(0)}\Big]^{2}$ and $\boldsymbol{g}_{\s}^{(-)}(u)$ is the regular solution of
	\begin{equation}
		\boldsymbol{g}_{\s}^{(-)''}+(-\frac{1}{u}+\frac{F'}{F})\boldsymbol{g}_{\s}^{(-)'}-\frac{B^{2}}{4F}\boldsymbol{g}_{\s}^{(-)}=0\ ,\quad \s=1,2\ .
	\end{equation}
	Since $\boldsymbol{g}_{\s}^{(-)}(u)$ and $\boldsymbol{g}^{1(-)}{}_2(u)$ that appears in the shear channel satisfy the same equations of motion and boundary conditions, we deduce that $\mathcal{G}_{\s}=\mathcal{G}_{12}$.
	
	Substituting \eqref{Gi} into the holographic energy-momentum tensor \eqref{BulkHT}, we find
	\begin{align}
		\delta T_{G_{\s}} =\Big(1+\frac{i\omega\mathcal{G}_{\s}}{2\kappa^{2}}u_{+}^{-\frac{3}{2}}\Big)\boldsymbol{g}_{\s}^{(-)}(0)+\mathcal{O}(\omega^{2})\ .
		\label{dtg12}
	\end{align}
	By definition, the diagonal components of the perturbed energy-momentum tensor can be expressed in terms of
	$\delta T_{G_{1}}, \delta T_{G_{2}}$ and $\delta T_{G_{+}}$ as
	\begin{align}
		\delta T_{11} & =\frac{1}{3}(2\delta T_{G_{1}}+\delta T_{G_{2}}+\delta T_{G_{+}})\ ,\nonumber \\
		\delta T_{22} & =\frac{1}{3}(-\delta T_{G_{1}}+\delta T_{G_{2}}+\delta T_{G_{+}})\ ,\nonumber \\
		\delta T_{33} & =\frac{1}{3}(-\delta T_{G_{1}}-2\delta T_{G_{2}}+\delta T_{G_{+}})\ .
	\end{align}
	Utilizing \eqref{dtg12} and the fact that up to first order in $\o$, $\delta T_{G_{+}}$ is real, we obtain
	\begin{align}
		{\rm Im}\delta T_{aa} & =\frac{\omega\mathcal{G}_{12}}{3\kappa^{2}}u_{+}^{-\frac{3}{2}}\sum_{b=1}^3M_{ab}\delta h_{bb},\quad M_{ab}=\left(\begin{array}{ccc}
			1 & -\frac{1}{2} & -\frac{1}{2}\\
			-\frac{1}{2} & 1 & -\frac{1}{2}\\
			-\frac{1}{2} & -\frac{1}{2} & 1
		\end{array}\right)\ ,
	\end{align}
	where  we have replaced $\mathcal{G}_{\s}$ by $\mathcal{G}_{12}$ and identified the leading FG coefficients of the perturbed metric via
	\be
	\boldsymbol{g}_{1}^{(-)}(0)=\delta h_{11}-\delta h_{22},\quad
	\boldsymbol{g}_{2}^{(-)}(0)=\delta h_{22}-\delta h_{33}\ .
	\ee
	Comparing to \eqref{dt}, we obtain the bulk viscosity as
	\begin{equation}
		\zeta_{ab}=\frac{2\mathcal{G}_{12}}{3\kappa^{2}} u_{+}^{-\frac{3}{2}} M_{ab}=\frac{s}{4\pi}\ \frac{4}{3}\mathcal{G}_{12} M_{ab}=\frac{4}{3}\eta_{(12)}M_{ab}\ .
		\label{zetaall}
	\end{equation}
	
	We recall that in terms of  the retarded Green's function, a conformal fluid has the property
	\be
	\lim_{\o\rightarrow 0}\frac{G^{TT,xy}_{xy}(\o)}{-\o}=\frac34\lim_{\o\rightarrow 0}\frac{G^{TT,xx}_{xx}(\o)}{-\o}=-\frac32\lim_{\o\rightarrow 0}\frac{G^{TT,yy}_{xx}(\o)}{-\o}\ .
	\ee
	According to our definition of the transport coefficients \eqref{KuboFormula}, the relations above imply precisely \eqref{zetaall}, which holds for any value of $B$. It is worth pointing out that
	in the previous work \cite{Grozdanov:2017kyl}, similar relations held only when $B\rightarrow 0$.


	\section{Conclusion and discussions} \label{section5}

	In this paper, we investigated a toy model of magnetohydrodynamics involving three dynamical U(1) gauge fields using holography. The gravity model comprises three 2-form gauge fields, which are dual to the three conserved currents of the magnetic 1-form symmetries associated with the three U(1) gauge fields. Different from previous work \cite{Hofman:2017vwr,Grozdanov:2017kyl}, this model admits an analytical AdS black brane solution in which the 2-form gauge fields carry equal fluxes  in mutually orthogonal directions. In the dual fluid description, these fluxes correspond to three mutually orthogonal background magnetic fields of equal magnitude, which implies that the SO(3) rotational symmetry in the spatial directions is broken down to the $S_3$ subgroup. Interestingly,  the symmetry of the gravitational  background solution suggests that this system admits a fortuitous SO(3) symmetry acting simultaneously on both the flavor index and the spatial indices of the 2-form gauge fields.   Owing to the existence of an analytical background solution, we are able to solve the perturbations of the 2-form gauge fields exactly, from which the resistivity coefficients are computed. In particular,
	we find that the resistivity coefficients $r_{[ab]}$ associated with the antisymmetric part of the dual 2-form current $\widetilde{J}_{[ab]}$ exhibit  novel features, depending on whether the external magnetic field is present or not. When the magnetic fields are turned off, $r_{[ab]}$ is simply identical to $r_{(ab)}$ and is inversely proportional to the temperature. However, once the magnetic fields are switched on, $r_{[ab]}$ differs significantly from $r_{(ab)}$ and becomes a discontinuous function of the temperature. The critical temperature at which the discontinuity occurs depends on both the magnetic field and the UV cutoff $\Lambda$ of the system. We also computed all the viscosity coefficients and found that all the shear viscosity coefficients are equal. For the bulk viscosity coefficients, the diagonal and off-diagonal components are each equal among themselves.  Additionally, we showed that the bulk viscosity is proportional to the shear viscosity.
	In particular,  $\zeta_{11} = \frac{4}{3}\eta$ and $\zeta_{12}= -\frac{2}{3}\eta$ for all values of the magnetic field. This work also yields an interesting byproduct: the ansatz for the energy-momentum tensor \eqref{t1} and the 2-form current \eqref{j1} can be applied to realistic hydrodynamic systems with anisotropy in all three spatial directions. Specifically, the vector field $h_\m^a$ can be replaced by any orthogonal basis spanning the spatial directions.
	
	As for future directions, we can generalize our analysis to other dimensions.  In fact, the gravity model that accommodates an analytical AdS magnetic black brane solution admits a generalization in $d+1$ dimensions \cite{Meiring:2023wwi}
	\be
	S=\int\sqrt{-g}d^{d+1}x\Big(R+\frac{d(d-1)}{2}-\frac14\sum_{i=1}^{d-1}\sum_{j=i+1}^{d-1}F^{ij}_{\m\n}F^{ij\m\n}\Big)\ .
	\ee
	Upon dualizing the 1-form gauge field to a $d-2$-form gauge field, the gravity model can describe
	$d$-dimensional plasma with $(d-1)(d-2)/2$ number of $d-2$-form conserved currents $J^{[ij]}_{(d-2)}=*dA_{(1)}^{[ij]}$ where $A^{[ij]}_{(1)}$ is the U(1) gauge field in the plasma. In the case of $d=3$, the dyonic black brane solution in the model above has been utilized to study dispersion relations of the hydro-modes \cite{Jeong:2022luo, Ahn:2022azl, Baggioli:2023oxa}. Clearly more aspects of the holographic magnetohydrodynamic system remains to be explored. It is also interesting to explore higher derivative corrections to the various transport coefficients. It would be particularly intriguing to see if these corrections can smooth out the discontinuity in $r_{[ab]}$.


	\section*{Acknowledgement}
	We are grateful to discussions with M. Baggiolli and L. Ma. This work is partially supported by National Natural Science Foundation of China (NSFC) under grants No. 12175164, No 12247103 and by the National Key Research and Development Program under grant No. 2022YFE0134300.

\end{document}